\documentstyle[12pt]{article}

\newcommand{\sect}[1]{\setcounter{equation}{0}\section{#1}}

\newcommand{\EQ}{\begin{equation}}
\newcommand{\EN}{\end{equation}}
\newcommand{\bea}{\begin{eqnarray}}
\newcommand{\ena}{\end{eqnarray}}
\newcommand{\vs}[1]{\vspace{#1 mm}}
\newcommand{\hs}[1]{\hspace{#1 mm}}
\def\bbox{{\,\lower0.9pt\vbox{\hrule \hbox{\vrule height 0.2 cm
\hskip 0.2 cm \vrule height 0.2 cm}\hrule}\,}}

\newcommand{\sumi}{\sum_{i=1}^{N}}
\newcommand{\sumij}{\sum_{i,j=1 \atop i<j}^{N}}
\newcommand{\suma}{\sum_{a=1}^{M-1}}
\newcommand{\waa}[1]{\sum_{#1>0}}

\renewcommand{\a}{\alpha}
\renewcommand{\b}{\beta}

\renewcommand{\c}{\gamma}
\renewcommand{\d}{\delta}
\newcommand{\e}{\epsilon}

\newcommand{\la}{\lambda}
\newcommand{\La}{\Lambda}

\newcommand{\pa}{\partial}

\newcommand{\dsl}{\pa \kern-0.5em /}
\newcommand{\nn}{\nonumber\\}

\newcommand{\shalf}{\frac{1}{2}}

\newcommand{\hi}{\langle}
\newcommand{\mi}{\rangle}

\begin{document}
\topmargin 0pt
\oddsidemargin 5mm

\newcommand{\NP}[1]{Nucl.\ Phys.\ {\bf #1}}
\newcommand{\AP}[1]{Ann.\ Phys.\ {\bf #1}}
\newcommand{\PL}[1]{Phys.\ Lett.\ {\bf #1}}
\newcommand{\NC}[1]{Nuovo Cimento {\bf #1}}
\newcommand{\CMP}[1]{Comm.\ Math.\ Phys.\ {\bf #1}}
\newcommand{\PR}[1]{Phys.\ Rev.\ {\bf #1}}
\newcommand{\PRC}[1]{Phys.\ Rep.\ {\bf #1}}
\newcommand{\PRL}[1]{Phys.\ Rev.\ Lett.\ {\bf #1}}
\newcommand{\PTP}[1]{Prog.\ Theor.\ Phys.\ {\bf #1}}
\newcommand{\PTPS}[1]{Prog.\ Theor.\ Phys.\ Suppl.\ {\bf #1}}
\newcommand{\MPL}[1]{Mod.\ Phys.\ Lett.\ {\bf #1}}
\newcommand{\IJMP}[1]{Int.\ Jour.\ Mod.\ Phys.\ {\bf #1}}
\newcommand{\JP}[1]{Jour.\ Phys.\ {\bf #1}}
\newcommand{\JMP}[1]{Jour.\ Math.\ Phys.\ {\bf #1}}

\begin{titlepage}
\setcounter{page}{0}
\begin{flushright}
OU-HET 245 \\
hep-th/9605083
\end{flushright}

\vs{10}
\begin{center}
{\Large
ALL THE EXACT SOLUTIONS OF GENERALIZED CALOGERO-SUTHERLAND MODELS
}
\vs{15}

{\large
Nobuyoshi Ohta\footnote{e-mail address: ohta@phys.wani.osaka-u.ac.jp}}

\vs{10}
{\em Department of Physics, Osaka University \\
Toyonaka, Osaka 560, Japan}

\end{center}
\vs{15}
\centerline{{\bf{Abstract}}}
\vs{5}

A collective field method is extended to obtain all the explicit solutions of
the generalized Calogero-Sutherland models that are characterized
by the roots of all the classical groups, including the solutions
corresponding to spinor representations for $B_N$ and $D_N$ cases.

\vs{10}

\end{titlepage}

\newpage

\sect{Introduction}

Recently there has been much interest in the Calogero-Sutherland (CS)
models,~\cite{CS,SUT} which describe one-dimensional many-body quantum
systems with inverse-square long-range interactions.
The reason for the interest is that the models play significant roles
in diverse subjects such as fractional statistics,~\cite{HAL,LM,POL,HA,LPS}
quantum Hall effect~\cite{K,AI,SF} and $W_\infty$ algebra.~\cite{HW}
The original models have a structure related to the classical group $A_N$.

Among many variants of the CS models,~\cite{HAL} a class of models have
been known to be exactly solvable and show interesting behaviors similar to
the original ones.~\cite{OP} In particular the so-called CS model of
$BC_N$-type (hereafter referred to as $BC_N$-CS model) is the most general
one with $N$ interacting particles. By setting various coupling constants
to zero, we can reduce the model to all other models of $B_N$, $C_N$ and
$D_N$ types. These models are known to be relevant to one-dimensional
physics with boundaries.

It was Stanley and Macdonald~\cite{SM} who found that the solutions for
$A_N$ type are expressed by Jack symmetric polynomials and studied their
properties. The explicit formulae for the wave functions for these models have
recently been obtained by the use of collective field method~\cite{JS,MP}
and conformal field theory technique by Awata et al.~\cite{AMOS,MY}
They showed that the Hamiltonian can be expressed in terms of Virasoro
and $W_M$ generators of positive modes and hence the solutions can be
represented as $W_M$ singular vectors, whose explicit forms are given by
integral representations using free bosons. (Here the notation
is slightly changed from that in ref.~\cite{AMOS}; $M$ is an arbitrary integer
$(\geq 2)$ which characterizes the $W$ algebra used in the construction
of the solutions and is independent of the number of the particles $N$.)
Unfortunately the wave functions were not known for the generalized
$BC_N$-CS models except for the ground states,~\cite{OP,BPS} though the energy
eigenvalues have been obtained for both ground and excited
states.~\cite{BPS,Y} These wave functions are important for examining
various properties of the models, like correlation functions.

In a previous paper~\cite{KO}, we have
given a systematic method to construct the wave functions for excited
states in these generalized models by extending the collective field
method. The method was applied to those solutions common to the
CS models of all types of the classical groups.
These solutions are the only ones to the models characterized by the root
systems of the $C_N$ and $BC_N$ groups. However, it turns out that
they do not exhaust the whole solutions for the generalized systems;
there are additional
solutions in the models of $B_N$ and $D_N$ types corresponding
to the spinor representations in these groups. In this paper, we will
derive all these solutions, namely wave functions and energy eigenvalues,
explicitly by extending our previous results using the collective
field method and clarify the whole structure of the solutions.

The paper is organized as follows. In order to establish our notations
and conventions, we summarize in \S~2 the $BC_N$-CS models characterized
by the root systems of the classical groups and the free field realization
of the $W_M$ algebra which is necessary for our subsequent discussions.
In \S~3, our method is explained and applied to the $BC_N$- and
$C_N$-CS models. In \S~4, we discuss the solutions for the $B_N$-CS
models. In \S~5, the solutions for the $D_N$-CS models are obtained.
Section~6 is devoted to discussions.

\sect{Preliminaries}

In this section, we summarize our notations and conventions which will
be used in the following discussions.

\subsection{Generalized CS models}

It has been known for some time~\cite{OP} that there exist a class
of models that are characterized by the root systems of the classical
groups and are exactly solvable. The Hamiltonian is given by
\EQ
H_{GCS}=-\sumi\frac{1}{2}\frac{\pa^2}{\pa q_i^2}
+\shalf\left(\frac{\pi}{L}\right)^2 \sum_{\vec{\a}\in R_{+}}
\frac{\mu_\a(\mu_\a+2\mu_{2\a}-1)|\vec{\a}|^2}
{\sin^2\frac{\pi}{L}(\vec{\a}\cdot\vec{q})},
\label{gcs}
\EN
where $R_+$ stands for positive roots of the classical group
under consideration and the coupling constants $\mu_\a$ are equal
for the roots of the same length. The most general model
is the one with all the roots in $B_N$ and $C_N$ algebras.
This is the $BC_N$-CS model we are going to discuss.

We first introduce the variables
\EQ
x_j\equiv\exp\left(\frac{2\pi i q_j}{L}\right); \qquad
D_i\equiv x_i\frac{\pa}{\pa x_i}.
\label{var}
\EN
Using these variables, the Hamiltonian (\ref{gcs}) is cast into
\bea
H_{GCS}(x_i;\b,\c,\d)&=& \shalf \left(\frac{2\pi}{L}\right)^2 \left[
 \sumi D_i^2-2\b(\b-1)\sumij  \left( \frac{x_i x_j}{(x_i-x_j)^2}
 +\frac{x_i x_j^{-1}}{(x_i-x_j^{-1})^2} \right) \right. \nn
& & \left. -\sumi  \left( \c(\c+2\d-1) \frac{x_i}{(x_i-1)^2}+4\d(\d-1)
 \frac{x_i^2}{(x_i^2-1)^2} \right) \right],
\label{hgcs}
\ena
where we have used $\b,\c,\d$ for coupling constants.
We note that putting $\c=0$ reduces the model to $C_N$-type,
$\d=0$ to $B_N$-type, and finally $\c=\d=0$ to $D_N$-type.
We also refer to these models as $C_N$-, $B_N$- and $D_N$-CS models,
respectively.

The ground state wave function and energy are given by~\cite{OP,BPS}
\bea
\Delta_{GCS} &=& \prod_{i=1}^{N} 
\left(\sin\frac{\pi}{L}q_i \right)^\c
\left(\sin\frac{2\pi}{L}q_i \right)^\d
\prod_{i,j=1 \atop i<j}^{N}
\left(\sin\frac{\pi}{L}(q_i-q_j)
\sin\frac{\pi}{L}(q_i+q_j) \right)^\b \nn
&\simeq& \prod_{i=1}^{N} x_i^{-\b(N-1)-\c/2-\d}
 (x_i-1)^\c (x_i^2-1)^\d
 \prod_{i,j=1 \atop i<j}^{N} (x_i-x_j)^\b(x_ix_j-1)^\b \nn
E_0^{GCS} &=& \sumi \left[ \frac{\c}{2} + \d+\b(N-i)\right]^2.
\label{grou}
\ena

Now our eigenvalue problem
\EQ
H_{GCS} \Delta_{GCS} \Phi^{GCS} = E_{GCS} \Delta_{GCS} \Phi^{GCS},
\EN
reduces to
\bea
&& H_{eff}\Phi^{GCS} = E_{eff} \Phi^{GCS} \;\; ; \nn
&& E_{GCS} = \shalf \left(\frac{2\pi}{L}\right)^2
 \left[ E_0^{GCS} + E_{eff} \right],
\label{eig}
\ena
where the effective Hamiltonian $H_{eff}$ acting on the function
$\Phi^{GCS}(x)$ is given by~\cite{KO}
\bea
H_{eff}(x_i:\b,\c,\d)&=& \sumi D^2_i + \b \sumij \left(
 \frac{x_i+x_j}{x_i-x_j}
 (D_i-D_j) + \frac{x_i+x_j^{-1}}{x_i-x_j^{-1}}(D_i+D_j) \right) \nn
&& +\sumi \left( \c \frac{x_i+1}{x_i-1}
 +2\d  \frac{x_i+x_i^{-1}}{x_i-x_i^{-1}}  \right)D_i .
\label{ham}
\ena
We express this Hamiltonian by free bosons and relate it to the
free boson representation of the $W_M$ algebra.
For this purpose, let us next summarize relevant results in the
free boson representation of this algebra. Since this is described
in detail elsewhere,~\cite{FL,AMOS,KO} we will be very brief.

\subsection{$W_M$ algebra}

Let $\vec{e}_i$ $(i=1,\cdots, M)$ stand for an orthonormal basis ($\vec{e}_i
\cdot \vec{e}_j = \d_{ij}$) for $A_M$ algebra. We define the weights of the
vector representation $\vec{h}_i$, the simple roots $\vec{\a}^a$ ($a=1,
\cdots, M-1$) and the fundamental weights $\vec{\La}_a$ by
\bea
&& \vec{h}_i = \vec{e}_i - \frac{1}{M}\sum_{j=1}^M \vec{e}_j, \qquad
\vec{\a}^a = \vec{h}_a-\vec{h}_{a+1}, \qquad
\vec{\La}_a = \sum_{i=1}^a \vec{h}_i, \nn
&& \vec{\a}^a\cdot \vec{\a}^b \equiv A^{ab} = 2\d^{a,b} - \d^{a,b+1}
-\d^{a,b-1}, \qquad
\vec{\a}^a\cdot \vec{\La}_b \equiv A^a_b = \d^a_b.
\ena
We then introduce $M-1$ free bosons
\EQ
\vec{\phi}(z)=\sum_{a=1}^{M-1}\phi^a(z)\vec{\La}_a
 =\sum_{a=1}^{M-1}\phi_a(z)\vec{\a}^a .
\EN
They have the mode expansion
\EQ
\vec{\phi}(z) = \vec{q} + \vec{a}_0 \ln z - \sum_{n \neq 0} \frac{1}{n}
\vec{a}_n z^{-n},
\EN
with the commutation relations
\EQ
[a_n^a,a_m^b]=A^{ab} n\d_{n+m,0}, \qquad
[a_0^a,q^b]= A^{ab},
\EN
The boson Fock space is generated by the oscillators of
negative modes on the highest weight state
\EQ
|\vec{\la} \mi = e^{\vec{\la}\cdot\vec{q}} |\vec{0}\mi; \qquad
\vec{a}_n |\vec{0}\mi = 0 \;\; (n\geq 0).
\EN
We define $\hi \vec{\la} |$ similarly with the inner product
$\hi\vec{\la} |\vec{\la}' \mi=\d_{\vec{\la},\vec{\la}'}$.

We need only the spin 2 and 3 generators of the $W_M$ algebra, which
are given by~\cite{FL}
\bea
T(z) &\equiv& \sum_n L_n z^{-n-2} \nn
&=& \shalf (\pa\vec{\phi}(z))^2
 + \a_0 \vec{\rho}\cdot\pa^2 \vec{\phi}, \nn
W(z) &\equiv& \sum_n W_n z^{-n-3} \nn
&=& \sum_{a=1}^{M-1} (\pa\phi_a(z))^2 \left( \pa\phi_{a+1}(z)
 - \pa\phi_{a-1}(z) \right) \nn
&& + \a_0 \sum_{a,b=1}^{M-1} (1-a)A^{ab} \pa\phi_a(z)\pa^2\phi_b(z)
 + \a_0^2 \sum_{a=1}^{M-1} (1-a)\pa^3\phi_a(z),
\ena
where
\bea
\a_0 &=& \sqrt{\b}-\frac{1}{\sqrt{\b}}, \nn
\vec{\rho} &=& \sum_{a=1}^{M-1} \vec{\La}_a ; \;\;
(\vec{\rho})^2=\frac{1}{12}M(M^2-1).
\ena
The highest weight states of the $W_M$ algebra are created from
the vacuum by the vertex operator as
$|\vec{\la}\mi =:e^{\vec{\la}\cdot\vec{\phi}(0)}:|\vec{0}\mi$,
whose conformal weight $h(\vec{\la})$ and $W_0$-eigenvalue
$w(\vec{\la})$ are
\bea
h(\vec{\la}) &=& \shalf\left[ (\vec{\la}-\a_0 \vec{\rho})^2
 -\a_0^2(\vec{\rho})^2 \right], \nn
w(\vec{\la}) &=& \sum_{a=1}^{M-1} \left[
 \la_a^2(\la_{a+1} -\la_{a-1}) + \a_0 \left( 2(a-1)\la_a \right.\right.\nn
&& \left. \left. + (1-2a)\la_{a+1}\right) \la_a 
 + 2\a_0^2(1-a)\la_a \right].
\label{wei1}
\ena
Another formula which will be useful is
\EQ
h(\vec{\la}_{\vec{r},\vec{s}}^{+}-\sqrt{\b}\sum_{a=1}^{M-1} r^a\vec{\a}^a)
= h(\vec{\la}_{\vec{r},\vec{s}}^{+}) + \sum_{a=1}^{M-1} r^a s^a.
\label{wei2}
\EN

We can define singular vectors $|{\vec r},{\vec s}\mi $ at
level $\sum_{a=1}^{M-1}r^a s^a$ with the highest
weight $|\vec{\la}^\pm_{\vec{r},\vec{s}}\mi$, where
$\vec{\la}^\pm_{\vec{r},\vec{s}}$ is defined by
\bea
\vec{\la}^+_{\vec{r},\vec{s}} &=& \sum_{a=1}^{M-1}\left[
 (1+r^a-r^{a-1})\sqrt{\b}-(1+s^a)/\sqrt{\b}\right] \vec{\La}_a, \nn
\vec{\la}^-_{\vec{r},\vec{s}} &=& \sum_{a=1}^{M-1}\left[
 (1+r^a)\sqrt{\b}-(1+s^a-s^{a-1})/\sqrt{\b}\right] \vec{\La}_a.
\ena
The explicit forms of the singular vectors are known in an integral form
using free bosons, but will not be needed in our following discussions.
Suffice it to say that they are annihilated by Virasoro $L_n$ and $W_n$
generators of positive modes and correspond to the following Young
diagrams parameterized by the numbers of boxes in each
row, $\la=(\la_1,\cdots,\la_M)$, $\la_1\geq\cdots\geq\la_M\geq 0$:

\vs{2}
\noindent
\makebox[ .2cm]{ }
\makebox[  2cm]{$s^1$}\hskip-.4pt
\makebox[1.7cm]{$s^2$}
\makebox[1.7cm]{ }
\makebox[1.4cm]{$s^{M-2}$}\hskip-.3pt
\makebox[1.3cm]{$s^{M-1}$}
\hfill\break
\makebox[  .2cm][r]{$\hfill\la=$}
\framebox[  2cm][l]{\rule[  -1cm]{0cm}{  2cm}$r^1$}\hskip-.4pt
\framebox[1.7cm][l]{\rule[-0.7cm]{0cm}{1.7cm}$r^2$}
\makebox[1.4cm]   {\raisebox{.25cm}{$\cdots\cdots$}}
\framebox[1.4cm][l]{\rule[-0.4cm]{0cm}{1.4cm}\raisebox{.25cm}{$r^{M-2}$}
     }\hskip-.4pt
\framebox[1.3cm][l]{\rule[-0.2cm]{0cm}{1.2cm}\raisebox{.25cm}{$r^{M-1}$}}
\makebox[.5cm][r]{.}
\vs{3}

\noindent
We can read off the relation between $\la$ and $\vec{r},\vec{s}$ from
this diagram.

\sect{Exact solutions for $BC_N$- and $C_N$-CS models}

In this section, we begin with the brief description of the exact
solutions for the models of $BC_N (\c \neq 0)$ and $C_N (\c=0)$ types
with $\d \neq 0$. These are also common solutions to all the CS models
if we set the coupling constants to zero appropriately. This was
the main result in our previous paper,~\cite{KO} but will be heavily used
in our following construction of all exact solutions.

First note that our system~(\ref{ham}) has the
reflection invariance under $x_i \to x_i^{-1}$ for each $i$ in addition
to the permutation symmetry under $x_i \leftrightarrow x_j$.
In fact, the solutions can be given by the symmetric power sums
\EQ
p_n=\sumi (x_i^n+x_i^{-n}).
\label{power}
\EN
It is known in mathematical literature~\cite{YOK} that the representation
ring for $BC_N$ and $C_N$ systems is isomorphic to the ring generated by
these symmetric power sums. Hence all the solutions for these systems,
which correspond to representations in the algebras, can
be obtained by using these functions (\ref{power}).

In terms of (\ref{power}), we can express our effective Hamiltonian.
This Hamiltonian is then mapped into oscillator representation by
\bea
| f\mi\mapsto f(x)&\equiv&\hi\vec{\la}| C_{\b}| f\mi \nn
C_{\b} &\equiv& \exp \left(\sqrt{\b}\waa{n}
\frac{1}{n}a_{n,1}p_n\right),
\label{map}
\ena
which gives the following correspondence between the oscillators
and the power sums (\ref{power}):
\EQ
\sqrt{\b} p_n \leftrightarrow a_{-n}^1 ;\;\;
\frac{n}{\sqrt{\b}}\frac{\pa}{\pa p_n} \leftrightarrow a_{n,1} .
\label{rule}
\EN
The Hamiltonian is further rewritten using the Virasoro generator $L_n$
of positive modes and $W_0$. The result of this series of manipulations
is~\cite{KO}
\bea
\hat{H}_{eff} &=& \hat{H}' + \sum_{n>0} \hat{H}_{n}
 + \sum_{a>1}\waa{n} a_{-n}^a (\cdots) \nn
&& + \sqrt{\b}\waa{n}\left(\frac{2}{N}a_{-n}^1 L_n-2a_{n,1}L_n\right)
+2\waa{n}\left\{\c L_n+\left(2 \d-\b\right)L_{2n}\right\},
\label{res}
\ena
where
\bea
\hat{H}' &=& \waa{n}\vec{a}_{-n}\cdot\vec{a}_n \left( 2N \b-1+\c+2\d
 - 2\sqrt{\b}a_{0,1} \right) +\sqrt{\b}\left(W_0-W_{0,zero}\right), \nn
\hat{H}_{n} &=& 2\c \suma \sum_{m=1}^{n-1}a_{n-m,a}
 \left( a_{m,a+1}-a_{m,a} \right)+\frac{2\c}{\sqrt{\b}} a_{n,1}
 \left\{ (n+1)(\b-1)+ N \b-\sqrt{\b}a_0^1 \right\} \nn
&&+\frac{2\c}{\sqrt{\b}}\sum_{a=2}^{M-1}a_{n,a}
 \left\{ (n+1)(\b-1)-\sqrt{\b}a_0^a \right\} \nn
&& + 2\sqrt{\b}\suma \sum_{m=1}^{n-1}
 a_{n,1}( a_{n-m,a}a_{m,a}-a_{n-m,a+1}a_{m,a} ) \nn
&& + 2(\b-2\d)\suma \sum_{m=1}^{2n-1}a_{2n-m,a}(a_{m,a}-a_{m,a+1}) \nn
&& - 2 (a_{n,1})^2 \left\{ (n+1)(\b-1)+N \b -\sqrt{\b}a_0^1 \right\} \nn
&&-2\sum_{a=2}^{M-1} a_{n,1}a_{n,a}
 \left\{ (n+1)(\b-1)-\sqrt{\b}a_0^a \right\} \nn
&&-\frac{2}{\sqrt{\b}} \left[ (\b-2\d)a_{2n,1} \left\{
 (2n+1)(\b-1)+N \b-\sqrt{\b}a_0^1 \right\} +\b a_{2n,1} \right]\nn
&&-\frac{2}{\sqrt{\b}}(\b-2\d)\sum_{a=2}^{M-1}a_{2n,a}
 \left\{ (2n+1)(\b-1)-\sqrt{\b}a_0^a \right\}.
\label{res1}
\ena
Here caret on the Hamiltonian implies that it is expressed in terms of
oscillators and $W_{0,zero}$ in $\hat{H}'$ is the zero mode part of $W_0$.
The third term involving $a_{-n}^a \; (a>1,n>0)$ in (\ref{res})
vanishes after multiplying by $\hi \vec{\la}|C_{\b}$ and may be
disregarded in the following.
An important observation is that $\hat{H}'$ is the sum of number
operators and $W_M$ zero mode and also that $\hat{H}_{n}$ consist of
annihilation operators only. It is at this point that the Virasoro
generators $L_n$ and $W_0$ are necessary; they are used to put cubic
terms involving $a_{-n}^1 (n > 0)$ in (\ref{res}) into the form
combined with $L_n$.

To construct our eigenstates of the Hamiltonian $\hat{H}_{eff}$, we take
singular vectors at the level $\sum_{a=1}^{M-1}r^a s^a$. Since these are
annihilated by Virasoro generators $L_n$ of positive modes,
only the first two terms in (\ref{res}) are relevant to our problem.
These are already eigenstates of $\hat{H}'$ with the eigenvalue~\cite{KO}
\bea
E_{\la}&=& \left[ h\left(\vec{\la}^+_{\vec{r},\vec{s}}
 - \sqrt{\b}\sum_{a=1}^{M-1}r^a\vec{\a}^a\right)
 - h\left(\vec{\la}^+_{\vec{r},\vec{s}} \right)\right] \nn
&& \times \left[ 2N \b-1+\c+2\d - 2 \left( \b r_1-s_1
+\sqrt{\b}\a_0\rho_1\right) \right] \nn
&& + \sqrt{\b} \left[ w\left(\vec{\la}^+_{\vec{r},\vec{s}}
 - \sqrt{\b}\sum_{a=1}^{M-1}r^a\vec{\a}^a\right) -
w\left(\vec{\la}^+_{\vec{r},\vec{s}} \right) \right] \nn
&=& \sum_{a=1}^{M-1} r^a s^a s^a+2\sum_{a,b=1\atop a>b}^{M-1}r^a s^a s^b \nn
&& +\sum_{a=1}^{M-1} r^a s^a (2N \b-\b+\c+2\d-\b r^a), \nn
&=& \sumi \left[\la_i^2+2  \left\{ \b(N-i)+\frac{\c}{2}+\d
 \right\}\la_i \right].
\label{exen}
\ena
Here use has been made of eqs.~(\ref{wei1}) and (\ref{wei2}) in deriving
the second equality, and of the relation between $\la$ and $\vec{r},\vec{s}$
obtained from the Young diagram in getting the third equality.

It is noted in ref.~\cite{KO} that applying $\hat{H}_{n}$ on the singular
vectors produces only states at the lower levels, and that the excitation
energy is given by the eigenvalue given in (\ref{exen}); $E_{eff}=E_\la$.
Thus the eigenstates of our system can be written as
\EQ
|\Phi_{\la}^{GCS}\mi = |J_{\la}\mi + \sum_{\mu<\la}C_{\mu}|J_{\mu}\mi,
\label{wvf}
\EN
where $|J_\la\mi$ is the oscillator representation of the Jack polynomials
for the $A_N$ case (or the $W_M$ singular vectors) with the coefficients
$C_\mu$ to be determined from the highest state $|J_\la\mi$ by the
application of $\hat{H}_n$:
\bea
&& \hi J_{\nu}|\waa{n}\hat{H}_{n}|J_{\la}\mi
+\sum_{\mu<\la}C_{\mu}\hi J_{\nu}|\waa{n}\hat{H}_{n}|J_{\mu}\mi \nn
&& \hs{5}= C_{\nu} \left( E_{\la}-E_{\nu} \right), \hs{5}
\left( \mbox{}\nu<\la \right),
\label{mas}
\ena
The inner products are easily evaluated by using the oscillator algebra.
A systematic algorithm was given in ref.~\cite{KO} how to solve this master
equation (\ref{mas}) to determine the coefficients $C_\mu$ succesively starting
from $|J_\la\mi$. In this way the oscillator representation for our
system can be determined from the exact solution for $A_N$ case, which
are given by singular vectors of the $W_M$ algebra (but modified to
be reflection invariant).

The actual eigenstates in terms of the symmetric power sums
(\ref{power}) can then be read off from the explicit expression in
terms of the boson oscillators by the rule~(\ref{rule}).
For our later convenience, let us denote the solutions thus constructed
as
\EQ
\Phi^{GCS}_\la(x_i;\b,\c,\d).
\label{solbc}
\EN
The total energy is obtained from eqs. (\ref{grou}) and (\ref{exen})
as~\cite{BPS}
\EQ
E_0^{GCS}+E_{\la}=\sumi \left[\la_i+\b(N-i)+\frac{\c}{2}+\d\right]^2 .
\EN
A simple example of the solutions is
\EQ
p_1+\frac{2N\c}{2\b (N-1)+\c+2\d+1},
\label{exam}
\EN
with the excitation energy $E_\la=2\b (N-1)+\c+2\d+1$.

These general solutions will play important roles in our construction
of all solutions for other systems.

\sect{Exact solutions for $B_N$-CS models $(\d=0)$}

In this section, we will derive exact solutions for $B_N$-CS models.
The solutions described in the preceeding section are, of
course, solutions for the $B_N$-CS models if we put $\d=0$.
However, in addition to these solutions, it turns out that there are
other solutions corresponding to the spinor representations
for $B_N$-CS models.

Now the Hamiltonian in question is
\EQ
H_B^{eff}(x_i;\b,\c) = H^{eff}(x_i;\b,\c,\d=0).
\label{hamb}
\EN
It has been known~\cite{YOK} that the representation ring for
the $B_N$-CS model is isomorphic to the ring generated by (\ref{power})
{\em and}
\EQ
\Delta_B \equiv \prod_{i=1}^{N} \left( \sqrt{x_i} + \frac{1}{\sqrt{x_i}}
\right).
\EN
This is like a ``spin field'' in string theory and produces solutions
corresponding to spinor representations. Obviously the solutions
involving odd powers of $\Delta_B$ do not mix with those without $\Delta_B$.
Since the square of this function can be expressed in terms of the power
sums (\ref{power}), the solutions are divided into two different classes.

\vs{5}
\underline{$(i)$ Solutions of first class}

The first class is those that can be expressed solely by power sums
(\ref{power}). These are the solutions $\Phi^{GCS}_\la(x_i;\b,\c,0)$
already given in (\ref{solbc}) and the energy eigenvalues are given by
\EQ
\sumi \left[\la_i+\b(N-i)+\frac{\c}{2} \right]^2 .
\label{ene1}
\EN

\vs{5}
\underline{$(ii)$ Solutions of second class}

The second class is those that contain one $\Delta_B$.
We will refer to this class of solutions as spinor solutions.
We now show how to derive these spinor solutions in our approach.

First by applying our effective Hamiltonian (\ref{hamb}), we see
that $\Delta_B$ is an eigenstate:
\bea
H^{eff}_B(x_i;\b,\c) \Delta_B &=& E_{0,spinor}^B \Delta_B, \nn
E_{0,spinor}^B &=& \frac{N}{2} \left[ \shalf +\b(N-1)+\c \right].
\ena
Our new solutions are expressed as
\EQ
\Phi^{GCS}_B = \Delta_B \Psi_B^{spinor}.
\label{solb}
\EN
The effective Hamiltonian acting on the wave function $\Psi_B^{spinor}$
is then derived as
\bea
H^{eff}_{spinor}(x_i;\b,\c)  &\equiv& (\Delta_B)^{-1} H^{eff}_B
(x_i;\b,\c) \Delta_B - E_{0,spinor}^B \nn
&=& H^{eff}_B(x_i;\b,\c) + \sumi \frac{x_i-1}{x_i+1} D_i \nn
&=& H^{eff}(x_i;\b,\c-1,\d=1).
\ena
We thus see that our problem is reduced to the eigenvalue problem
for the first class of solutions with the value of $\c$ and $\d$
shifted to $\c-1$ and 1, respectively. Namely our problem goes back to
the solutions of $BC_N$-type with nonzero $\d$. Fortunately
this is already solved in the previous section, and hence
our solution is given by (\ref{solb}) with
\EQ
\Psi_B^{spinor} = \Phi^{GCS}_\la(x_i;\b,\c-1,1),
\label{spinsolb}
\EN
in terms of the solution in (\ref{solbc}).
The eigenvalues for these solutions are given by
\bea
E &=& E_0^{GCS}|_{\d=0} + E_{0,spinor}^B + E_\la|_{\c\to\c-1,\d=1} \nn
&=& \sumi \left[\la_i+\b(N-i)+\frac{\c}{2}+\shalf \right]^2 .
\label{enebs}
\ena
Thus the eigenenergy for this case is obtained from that (\ref{ene1}) of
the solution of the first class just by shifting $\la_i$ by $\shalf$.
This is the reflection of the fact that this second class of solutions
correspond to spinor representation of the classical groups $B_N$.

A simple example of this class of solutions is obtained from (\ref{exam}) as
\EQ
\left( p_1+\frac{2N(\c-1)}{2\b (N-1)+\c+2}\right) \Delta_B.
\EN

\sect{Exact solutions for $D_N$-CS models ($\c=\d=0$)}

We now turn to the solutions for $D_N$-CS models.
There are several complications in this model.
Some of the solutions for $D_N$ case are very similar to those for $B_N$.
The difference arises because there are two distinct classes of spinor
representations for $D_N$, and hence two classes of solutions corresponding
to these in addition to the solutions in \S~3. Not only those, we also
have further additional solutions.

Our Hamiltonian is
\EQ
H_D^{eff}(x_i;\b) = H^{eff}(x_i;\b,\c=0,\d=0).
\label{hamd}
\EN
The representation ring for the $D_N$-CS model is isomorphic~\cite{YOK}
to the ring generated by (\ref{power}) {\em and}
\EQ
\Delta^\pm = \prod_{i=1}^{N} \left(\sqrt{x_i} \pm \frac{1}{\sqrt{x_i}}
\right).
\EN
In a different basis
\bea
\Delta^+ \pm \Delta^- &=& \sum_{\e_1 \e_2 \cdots \e_N=\pm 1}
2 x_1^{\e_1/2}x_2^{\e_2/2} \cdots x_N^{\e_N/2} , \nn
&& (\e_i = \pm 1 ; \; i=1,\cdots, N),
\ena
we see that these correspond to the spinor representations of
opposite chirality.

Again the solutions involving odd powers of $\Delta^\pm$ do not mix with
the solutions without $\Delta^\pm$. Since the square of these functions
can be expressed in terms of the power sums (\ref{power}), our exact
solutions fall into several different classes.

\vs{5}
\underline{$(i)$ Solutions of first class}

The first class is again those that can be expressed solely by power
sums (\ref{power}). These are the solutions with $\c=\d=0$ already given
in \S~3:
\EQ
\Phi^{GCS}_\la(x_i;\b,0,0) ; \;\;
E = \sumi \left[ \la_i+\b(N-i)\right]^2.
\EN

\vs{5}
\underline{$(ii)$ Solutions of second class}

The second class is those that contain $\Delta^\pm$. This class is
further divided into three classes since the solutions involving $\Delta^\pm$
do not mix with each other and hence they are divided into those with
$\Delta^\pm$ and the product of these. (Even powers of these
functions are not independent and do not produce new solutions.)
These can be constructed
in exactly the same way as the spinor solutions for $B_N$.

First by applying our effective Hamiltonian (\ref{hamd}),
we see that $\Delta^\pm$ is an eigenstate with the same eigenvalue:
\bea
H^{eff}_D \Delta^\pm &=& E^D_{0,spinor} \Delta^\pm, \nn
E^D_{0,spinor} &=& \frac{N}{2} \left[ \shalf +\b(N-1) \right].
\ena
Our new solutions corresponding to spinor representations are expressed as
\EQ
\Phi^{GCS,\pm}_D = \Delta^\pm \Psi_{D,spinor}^\pm.
\label{spinsold}
\EN
The effective Hamiltonian acting on the wave function $\Psi_{D,spinor}^\pm$
is then derived as
\bea
H^{eff}_{D,\pm}  &\equiv& (\Delta^{\pm})^{-1} H^{eff}_D \Delta^\pm
 - E^D_{0,spinor} \nn
&=& H^{eff}_D + \sumi \frac{x_i\mp 1}{x_i\pm 1} D_i.
\ena
We thus see that our problem is again reduced to the eigenvalue problems
for the first class of solutions, with the value of $\c$ and $\d$
shifted as
\bea
&& \left\{ \begin{array}{l}
\c=0 \to -1 \\
\d=0 \to 1  \end{array}  \right. {\rm for} \;\; \Delta^+ , \nn
&& \left\{ \begin{array}{l}
\c=0 \to 1 \\
\d=0 \to 0  \end{array}  \right. \;\;\; {\rm for} \;\; \Delta^-.
\ena
The first one $\Psi_{D,spinor}^+$ gives the same solution (\ref{spinsolb})
as the $B_N$ case with energy eigenvalues (\ref{enebs}), in which we
should put $\c=0$. The second one is also already solved in \S~3,
and hence the solutions are given by (\ref{spinsold}) with
\bea
\Psi_{D,spinor}^+ &=& \Phi_\la^{GCS}(x_i;\b,-1,1), \nn
\Psi_{D,spinor}^- &=& \Phi_\la^{GCS}(x_i;\b,1,0).
\ena
The eigenvalues for the solutions $\Phi_D^{GCS,-}$ are given by
\bea
E &=& E_0^{GCS}|_{\c=\d=0} + E_{0,spinor}^D + E_\la |_{\c=1,\d=0} \nn
&=& \sumi \left[\la_i+\b(N-i) + \shalf \right]^2 ,
\ena
which are degenerate with those for $\Phi_D^{GCS,+}$.
Thus the energy eigenvalues for this case are again obtained from those of
the solutions of the first class just by shifting $\la_i$ by $\shalf$.

Simple examples again from eq.~(\ref{exam}) are
\EQ
\left( p_1 \mp \frac{N}{\b (N-1)+1}\right) \Delta^\pm.
\EN
These examples of the solutions for $N=2$ were pointed out to the author
by D. Serban.

\vs{5}
\underline{$(iii)$ Solutions of third class}

The last class of solutions have the form
\EQ
\Phi^{GCS}_D = \Delta^+ \Delta^- \Psi_{D}.
\EN
Again $\Delta^+ \Delta^-$ is an eigenstate of $H^{eff}_D$ with
eigenvalue
\EQ
E^D_0 = N [ 1 +\b(N-1) ].
\EN

The effective Hamiltonian acting on the wave function $\Psi_D$
is then derived as
\bea
H^{eff}_D  &\equiv& (\Delta^+\Delta^-)^{-1} H^{eff}_D \Delta^+\Delta^-
 - E^D_{0} \nn
&=& H^{eff}_D + 2\sumi \frac{x_i+x_i^{-1}}{x_i-x_i^{-1}} D_i \nn
&=& H^{eff}(x_i;\b,\c=0,\d=1).
\ena
We thus see that our problem is again reduced to the eigenvalue
problems for the first class of solutions, with the value of $\d$
shifted as $\d=0 \to 1$, resulting in the $BC_N$ case.
The solutions are thus
\EQ
\Psi_D = \Phi_\la^{GCS}(x_i;\b,0,1).
\EN
The energy eigenvalues for these solutions are given by
\bea
E &=& E_0^{GCS}|_{\c=\d=0} + E_0^D + E_\la |_{\c=0,\d=1} \nn
&=& \sumi \left[\la_i+\b(N-i) + 1 \right]^2 .
\ena
Though these are similar to those for the first class of solutions
with the value of $\la_i$ shifted by 1, the symmetry of the wave
functions is different; they are anti-symmetric under the reflection.

This exhausts the exact solutions for the $D_N$-CS models.

\sect{Conclusions and discussions}

In this paper, we have first given a systematic algorithm to compute
eigenfunctions for excited states for the most general $BC_N$-CS models.
It is remarkable that these can be easily obtained from those for the $A_N$
case (but modified to be reflection invariant), which are nothing but singular
vectors of the $W_M$ algebra. These are the only solutions for the systems
of $BC_N$ type with nonzero coupling constants $\c$ and $\d$ and of $C_N$ type
with $\c=0,\d \neq 0$.

When the coupling constant $\d=0$, the model reduces to the $B_N$-CS model,
which has additional solutions corresponding to the unique spinor
representation of the group. If $\c$ is further set to zero, the model
reduces to $D_N$-CS models, which has three additional classes of solutions,
two of which correspond to the two distinct spinor representations of
the group. We have been able to derive all these solutions thanks to
the isomorphism between the `polynomial' ring and representations of
classical groups.~\cite{YOK}
These additional solutions can be actually expressed by using the universal
solutions described in \S~3 and a kind of ``spin fields''. In this sense,
the solutions discovered in ref.~\cite{KO} are the fundamental ones.

There are several interesting extensions of the present work. For example,
it should be straightforward to apply our method to CS models based on the
exceptional groups, which are also known to be exactly solvable.~\cite{OP}
Another immediate problem is whether this method can be applied to the
models involving elliptic functions.
It is also known~\cite{FMB} that there is a supersymmetric extension of
these models. The exact solvability of the extended models remains unchanged,
but this class of models may have deeper connection with the recent exact
solutions of four-dimensional $N=2$ super-Yang-Mills theory.~\cite{SW}

Another possible direction of investigation is the connection of the present
approach and the critical behavior of the models.~\cite{YKY} It was shown that
the critical behavior of these models are governed by $c=1$ conformal field
theory. It would be interesting to examine what is the relation between $W_M$
algebra in our formulation and the conformal field theory description
with $c=1$.

We hope that our present investigation motivates these studies and help to
shed some light on these problems.

\section*{Acknowledgments}

We would like to thank D. Serban for crucial comments on the
solutions corresponding to spinor representations which motivated
the present work. We would also like to thank K. Higashijima for
valuable discussions and indicating ref.~\cite{YOK}. Thanks are
also due to H. Awata, Y. Matsuo, S. Odake and T. Yamamoto for useful comments.


\newpage
\begin{thebibliography}{99}
\bibitem{CS} F. Calogero: \JMP{10} (1969) 2191, 2197; {\bf 12} (1971) 419.
\bibitem{SUT} B. Sutherland: \JMP{12} (1971) 246, 251;
 \PR{A4} (1971) 2019; {\bf A5} (1972) 1372.
\bibitem{HAL} N. Kawakami: \PTP{91} (1994) 189;\\
 F. D. M. Haldane, in the {\em Correlation Effects in Low Dimensional Electron
 Systems}, eds. A. Okiji and N. Kawakami (Springer, 1994);\\
 V. Pasquier: preprint SPhT/94-060, hep-th/9405104.
\bibitem{LM} J. M. Leinaas and J. Myrheim: \PR{B37} (1988) 9286.
\bibitem{POL} A. P. Polychronakos: \NP{B324} (1989) 597.
\bibitem{HA} Z. N. C. Ha: \PRL{73} (1994) 1574; \NP{B435} (1995) 604.
\bibitem{LPS} F. Lesage, V. Pasquier and D. Serban: \NP{B435} (1995) 585.
\bibitem{K} N. Kawakami: \PRL{71} (1993) 275.
\bibitem{AI} H. Azuma and S. Iso: \PL{B331} (1994) 107.
\bibitem{SF} M. Stone and M. Fisher: \IJMP{B8} (1994) 2539.
\bibitem{HW} K. Hikami and M. Wadati: \PRL{73} (1994) 1191.
\bibitem{OP} M. A. Olshanetsky and A. M. Perelomov: \PRC{71} (1981) 313;
 {\bf 94} (1983) 313.
\bibitem{SM} R. Stanley: Adv. Math. {\bf 77} (1989) 76;\\
 I. G. Macdonald: Lect. Note in Math. {\bf 1271} (Springer, 1987) p. 189.
\bibitem{JS} A. Jevicki and B. Sakita: \NP{B165} (1980) 511.
\bibitem{MP} J. A. Minahan and A. P. Polychronakos: \PR{B50} (1994) 4236.
\bibitem{AMOS} H. Awata, Y. Matsuo, S. Odake and J. Shiraishi:
 \PL{B347} (1994) 49; \NP{B449} (1995) 347.
\bibitem{MY} K. Mimachi and Y. Yamada: \CMP{174} (1995) 447.
\bibitem{BPS} D. Bernard, V. Pasquier and D. Serban: Europhys. Lett.
 {\bf 30} (1995) 301.
\bibitem{Y} T. Yamamoto: J. Phys. Soc. Japan {\bf 63} (1994) 1223.
\bibitem{KO} M. Kojima and N. Ohta: \NP{B473} (1996) 455.
\bibitem{FL} V. Fateev and S. Lykyanov: \IJMP{A3} (1988) 507.
\bibitem{YOK} I. Yokota: {\em Groups and Representaions}, (Syokabo,
 Tokyo, 1981).
\bibitem{FMB} D. Z. Freedman and P. F. Mende: \NP{B344} (1990) 317;\\
 L. Brink, T. H. Hansson, S. Konstein and M. A. Vasiliev: \NP{B401}
 (1993) 591.
\bibitem{SW} N. Seiberg and E. Witten: \NP{B426} (1994) 19.
\bibitem{YKY} T. Yamamoto, N. Kawakami and S.-K. Yang: preprint YITP/K-1117
 (1996).
\end{thebibliography}
\end{document}